\documentclass[9pt,twocolumn,twoside]{pnas-new}

\templatetype{pnasresearcharticle} 

\title{Scaling and Spontaneous Symmetry Restoring of Topological Defect Dynamics in Liquid Crystal}

\author[a]{Yohei Zushi}
\author[a,1]{Kazumasa A. Takeuchi} 

\affil[a]{Department of Physics, The University of Tokyo, Tokyo 113-0033, Japan}

\leadauthor{Zushi} 

\significancestatement{
When order is formed, topological defects may appear generically, and as such, they are a key to characterize and control the order formation observed in different fields of physics and related disciplines.
At the heart of such approaches is disentangling universal and non-universal aspects of topological defect dynamics.
Here we do so by realizing direct observation of 3D dynamics of liquid crystalline defects.
Analyzing topological rearrangements called reconnections, on the one hand we establish the validity of the scaling law known for quantum fluid and 2D liquid crystal; on the other hand, we reveal that the asymmetry present in 2D defect dynamics disappears in 3D.
This finding is accounted for by a general mechanism that may hold beyond liquid crystal.
}

\authorcontributions{
K.A.T. conceived the project and directed the research. Y.Z. prepared the experimental samples and carried out the experiments. Both authors analyzed the data and contributed to the interpretation of the results. Both authors wrote and edited the manuscript.
}
\authordeclaration{
The authors declare no competing interests.
}
\correspondingauthor{\textsuperscript{1}To whom correspondence should be addressed. E-mail: kat@kaztake.org}

\keywords{liquid crystal $|$ topological defect line $|$ reconnection $|$ scaling law} 

\begin{abstract}
Topological defects -- locations of local mismatch of order -- are a universal concept playing important roles in diverse systems studied in physics and beyond, including the Universe, various condensed matter systems, and recently, even life phenomena.
Among these, liquid crystal has been a platform for studying topological defects via visualization,
yet it has been a challenge to resolve three-dimensional structures of dynamically evolving singular topological defects.
Here we report a direct confocal observation of nematic liquid crystalline defect lines, called disclinations, relaxing from an electrically driven turbulent state.
We focus in particular on reconnections, characteristic of such line defects.
We find a scaling law for in-plane reconnection events, by which the distance between reconnecting disclinations decreases by the square root of time to the reconnection.
Moreover, we show that apparently asymmetric dynamics of reconnecting disclinations is actually symmetric in a co-moving frame, in marked contrast to the two-dimensional counterpart whose asymmetry is established.
We argue, with experimental supports, that this is because of energetically favorable symmetric twist configurations that disclinations take spontaneously, thanks to the topology that allows rotation of winding axis.
Our work illustrates a general mechanism of such spontaneous symmetry restoring that may apply beyond liquid crystal, which can take place if topologically distinct asymmetric defects in lower dimensions become homeomorphic in higher dimensions and if the symmetric intermediate is energetically favorable.
\end{abstract}

\dates{This manuscript was compiled on \today}
\doi{\url{www.pnas.org/cgi/doi/10.1073/pnas.2207349119}}

\usepackage{comment,physics}

\newcommand{\unit}[1]{~\mathrm{#1}}

\newcommand{\diff}[2]{\frac{\mathrm{d} #1}{\mathrm{d} #2}}

\renewcommand{\eqref}[1]{Eq.\,\ref{#1}}

\newcommand{\figref}[1]{Fig.\,\ref{#1}}

\newcommand{\supfigref}[1]{\textit{SI Appendix}, Fig.\,\ref{#1}}
\newcommand{\suptblref}[1]{\textit{SI Appendix}, Table\,\ref{#1}}
\newcommand{\supsecref}[1]{\textit{SI Appendix}}
\newcommand{\vidref}[1]{Movie\,#1}
\newcommand{\myetal}{\textit{et al}.\ }

\usepackage{xr} 
\makeatletter
\newcommand*{\addFileDependency}[1]{
  \typeout{(#1)}
  \@addtofilelist{#1}
  \IfFileExists{#1}{}{\typeout{No file #1.}}
}
\makeatother

\newcommand*{\myexternaldocument}[1]{
    \externaldocument[S-]{build/#1}
    \addFileDependency{#1.tex}
    \addFileDependency{build/#1.aux}
}
\myexternaldocument{suppl-arxiv}

\begin{document}

\maketitle
\thispagestyle{firststyle}
\ifthenelse{\boolean{shortarticle}}{\ifthenelse{\boolean{singlecolumn}}{\abscontentformatted}{\abscontent}}{}


\dropcap{T}opologically nontrivial configurations of order, called topological defects, may appear generically and spontaneously when order is formed.
As such, topological defects have been studied in diverse disciplines \cite{Nakahara-Book2003,Chaikin.Lubensky-Book2000}, including cosmology \cite{Vilenkin.Shellard-Book2000}, crystals and liquid crystals \cite{Chaikin.Lubensky-Book2000}, 
superconductivity and superfluid \cite{Zurek-PR1996,Vollhardt.Woelfle-book2013,Bewley.etal-PNAS2008,Fonda.etal-PNAS2014,Minowa.etal-SA2022,Serafini.etal-PRX2017}, 
and biology \cite{Saw.etal-N2017,Kawaguchi.etal-N2017,Doostmohammadi.etal-NC2018,Copar.etal-PRX2019,Binysh.etal-PRL2020,Duclos.etal-S2020,Copenhagen.etal-NP2020,Maroudas-Sacks.etal-NP2021,Ruske.Yeomans-PRX2021,Shimaya.Takeuchi-a2021}, to name but a few.
While there exist various kinds of defects characterized by different symmetries and properties, defects may also enjoy common properties across different disciplines.
In this context, liquid crystal has the advantage that it is amenable to direct optical observations, various compounds and techniques exist, and, as a soft matter system, it shows large response to external fields, being suitable for studying nonequilibrium and nonlinear effects \cite{Chaikin.Lubensky-Book2000,deGennes.Prost-Book1995}. 
This advantage has been recognized and used for decades, with a notable example of observing liquid crystal defects to test predictions for cosmic strings \cite{Chuang.etal-S1991}.
Moreover, the scope of studies of liquid crystalline defects has been recently extended remarkably, including the use of defects as templates for molecular self-assembly \cite{Wang.etal-Nmat2016} and the recent surge of investigations of active nematic systems bearing relevance to life phenomena \cite{Saw.etal-N2017,Kawaguchi.etal-N2017,Doostmohammadi.etal-NC2018,Copar.etal-PRX2019,Binysh.etal-PRL2020,Duclos.etal-S2020,Copenhagen.etal-NP2020,Maroudas-Sacks.etal-NP2021,Ruske.Yeomans-PRX2021,Shimaya.Takeuchi-a2021}.

Despite this history, resolving fully three-dimensional (3D) structures of liquid crystal defects has not been straightforward, even for the simplest kind of defects, namely nematic disclination lines.
Well-known techniques for 3D observation of defects and other orientational structures are the fluorescence confocal polarizing microscopy \cite{Smalyukh.etal-CPL2001,Lavrentovich-PJP2003} and two- or three-photon excitation fluorescence polarizing microscopy \cite{Lee.etal-OL2010, Trivedi.etal-PNAS2012, Ackerman.Smalyukh-PRE2016}.
Both techniques allow one to reconstruct the 3D structure of the director field, by which one can determine the position and structure of defects in principle.
To do so, however, one needs to reduce the effect of defocusing and polarization changes due to the birefringence of liquid crystal.
For singular defects such as nematic disclinations, scattering at the core gives another difficulty.
The effect of birefringence can be significantly reduced by partial polymerization of the medium
\cite{Evans.etal-PRE2013}, but this cannot be used to study dynamics of defects.

\begin{figure}[t]
\centering
\includegraphics[width=.92\hsize]{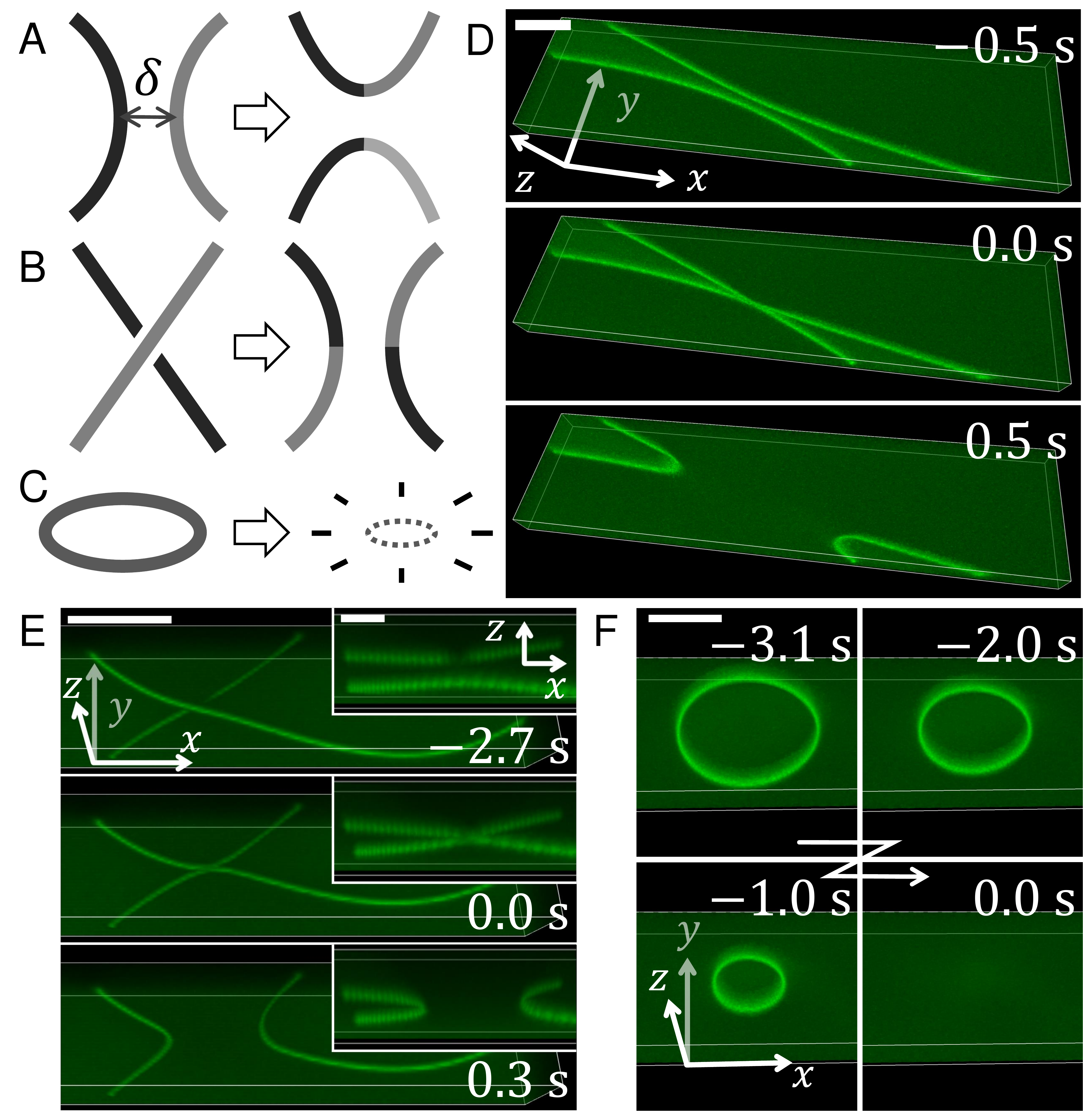}
\caption{Reconnections and loop shrinkage. (\textit{A-C}) Sketches of an in-plane reconnection (\textit{A}), an intersecting reconnection (\textit{B}), and a loop shrinkage (\textit{C}). (\textit{D-F}) Confocal observations of an in-plane reconnection (\textit{D}), an intersecting reconnection (\textit{E}), and a loop shrinkage (\textit{F}). The inset of (\textit{E}) displays a side view of the event shown in the main panel. The scale bars represent $50\unit{\mu{}m}$ for the main panels and $20\unit{\mu{}m}$ for the inset of (\textit{E}). See also Supplementary Videos~1-4.}
\label{fig1}
\end{figure}

Here we propose a method to capture dynamically evolving 3D structures of nematic disclination lines, by using confocal microscopy and a recently reported accumulation of fluorescent dyes around the singular core of defects \cite{Ohzono.etal-SR2016}.
This method allows us to visualize the disclinations directly (\figref{fig1}), without reconstructing and analyzing the director field.
Using this technique, we observe reconnections of disclinations -- a hallmark of such topological defect lines -- and characterize the reconnection dynamics in terms of scaling and symmetry.

\subsection*{Observations of disclination dynamics}
To study disclination dynamics, we add fluorescent dye to liquid crystal and observe the fluorescence from the dye localized at the 3D disclinations by confocal microscopy. Using the previously reported apparent length scale of dye accumulation, $\approx 0.33\unit{\mu m}$ \cite{Ohzono.etal-SR2016}, and the typical value of the diffusion coefficient of dye molecules, $\approx 10^{-10}\unit{m^2/s}$ \cite{Lavrentovich-PJP2003,Blinov.Chigrinov-Book1994}, we evaluate that dye can follow the evolution of disclinations at the time resolution of roughly $1\unit{ms}$. This is to compare with the time scale of the disclination dynamics, which can be evaluated at $\gamma_1\ell^2/K$ with Frank constant $K$, rotational viscosity $\gamma_1$, and characteristic length scale $\ell$ of disclination lines (such as the radius of curvature) \cite{deGennes.Prost-Book1995}. For typical mesogens (including the one used in this work), we have $K \approx 10^1\unit{pN}$ and $\gamma_1 \approx 10^2\unit{mPa\cdot s}$ \cite{deGennes.Prost-Book1995}, so that disclinations of length scale, e.g., $\ell \approx 10^1\unit{\mu m}$, evolve over a time scale of roughly $1\unit{s}$. Therefore, the disclination dynamics can be faithfully captured by confocal images acquired at a time interval between $1\unit{ms}$ and $1\unit{s}$ (or longer for disclinations of larger length scales). To fulfill this condition, we chose a laser-scanning confocal microscopy equipped with a resonant scanner working at $8\unit{kHz}$ and a piezo objective scanner (see Methods for details). 

A liquid crystal sample, MLC-2037 doped with fluorescent dye Coumarin 545T and electrolyte tetra-$n$-butylammonium bromide, was filled in a cell that consists of parallel glass plates with transparent electrodes (indium tin oxide) separated by $130\unit{\mu m}$ thick spacers (see Methods).
The planer alignment condition was imposed on the cell surfaces.
We generated a large density of disclinations by using an electrohydrodynamic turbulence \cite{deGennes.Prost-Book1995}, induced by an electric field applied to the liquid crystal sample.
The electric field was then removed and disclinations started to undergo relaxation.
We indeed observed a large density of singular disclinations upon removal of the electric field, followed by coarsening dynamics including reconnections and loop shrinkage (\figref{fig1} and Supplementary Videos~1-4), similar to those observed previously by bright-field microscopy \cite{Chuang.etal-S1991,Yurke.etal-PB1992,Chuang.etal-PRE1993,Mather.etal-LC1996,Ishikawa.Lavrentovich-EL1998}.
We also observed nonsingular disclinations terminating at singular ones (\supfigref{S-figS1}), as well as other kinds of defect structure, as reported in past bright-field studies \cite{Yurke.etal-PB1992,Chuang.etal-PRE1993}.

Most disclination lines were found near the midplane between the top and bottom surfaces and extended mostly horizontally, because of the homogeneous boundary condition we imposed.
As a result, most of the observed reconnections were classified into the following two kinds: in-plane reconnections (\figref{fig1}A,D and \vidref{2}) and intersecting reconnections (\figref{fig1}B,E and \vidref{3}).
An in-plane reconnection consists of a pair of curved disclinations in a nearly single horizontal plane, which approach in that plane and reconnect (\figref{fig1}D and \vidref{2}).
An intersecting reconnection consists of a pair of disclinations crossing at different $z$ positions, which approach vertically and reconnect (\figref{fig1}E and \vidref{3}).
In this case, the upper disclination appeared dark above the intersection (\figref{fig1}E inset) and apparently bent when the pair is close enough, presumably because of the lensing effect due to the lower disclination.
Since this prevented quantitative analysis, in the following we focus on the in-plane reconnections and study their reconnection dynamics.
We analyzed a total of $40$ in-plane reconnections without any noticeable nonsingular disclinations in the field of view.

\subsection*{Scaling law for in-plane reconnections}

\begin{figure}[t]
\includegraphics[width=\hsize,clip]{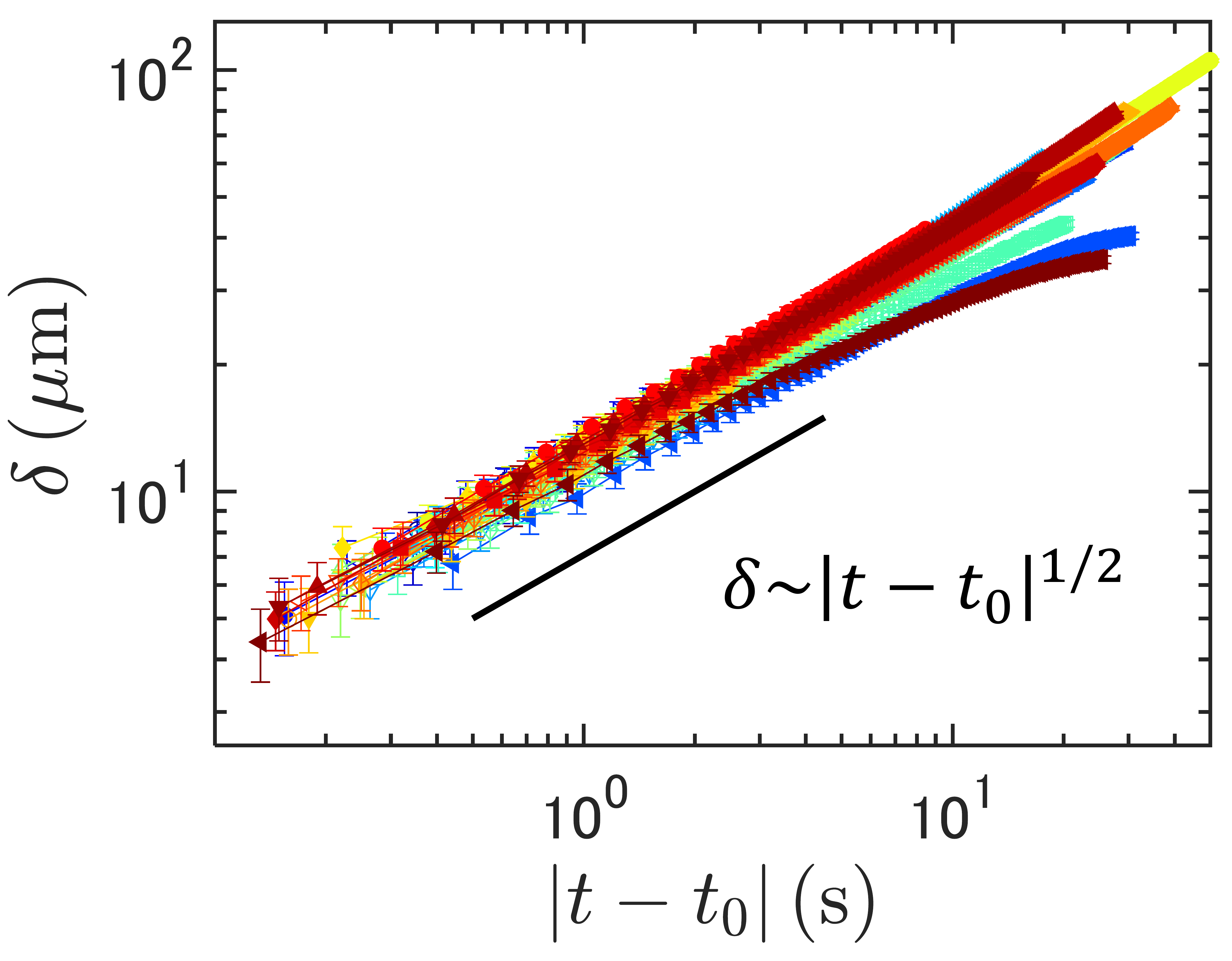}
\centering
\caption{Scaling law for in-plane reconnections. Results for all $40$ reconnection events are shown with different colors. The error bars indicate the uncertainty evaluated from the slicing and the Gaussian fitting used to determine the coordinates of the disclinations (see Methods).}
\label{fig2}
\end{figure}

Using the confocal images of the in-plane reconnections, we extracted the 3D positions of the two disclinations, until the moment of the reconnection (see Methods).
Measuring how the minimum distance between the two disclinations, $\delta(t)$, decreases with time $t$ (\figref{fig2}), we found the following scaling law for \textit{all} in-plane reconnections:
\begin{equation}
    \delta(t) \simeq C |t-t_0|^{1/2},  \label{eq:delta}
\end{equation}
with a coefficient $C$.
This power law is identical to that for annihilation of point disclinations in 2D nematics \cite{Svensek.Zumer-PRE2002,Denniston-PRB1996}, as well as that for reconnections of quantum vortices in quantum fluids \cite{Bewley.etal-PNAS2008,Fonda.etal-PNAS2014,Minowa.etal-SA2022}.
It is interesting to note that interaction of disclinations in 3D nematics was theoretically evaluated only very recently \cite{Long.etal-SM2021,Schimming.Vinals-SM2022}, and the power law in \eqref{eq:delta} was derived in the case of straight disclinations.
Although the experimentally observed disclinations were not straight but curved inward (\figref{fig1}D), we may argue that the time evolution of $\delta(t)$ is dominated by interaction between the two closest points, so that the disclination curvature did not affect the observed power law significantly.
The scaling law (\eqref{eq:delta}) was also observed numerically for curved disclinations in Ref.\cite{Schimming.Vinals-SM2022}.
Of course, it is important to extend those theoretical approaches to the case of curved disclinations and confirm the robustness of the power law in \eqref{eq:delta}.

\subsection*{Apparent asymmetry in the laboratory frame}
In the case of 2D nematics, disclinations are point-like and characterized by a topological invariant called the winding number.
Energetically stable are disclinations of winding number $\pm 1/2$ (see the left and right sketches in \figref{fig4}A below), and disclinations of opposite signs attract each other, approach, and annihilate. 
Here, it is well known that such a pair of $+1/2$ and $-1/2$ disclinations approaches in an asymmetric manner, due to the different backflow generated by the two defects \cite{Toth.etal-PRL2002,Svensek.Zumer-PRE2002}.
It would be then natural to expect analogous asymmetry to arise for line disclinations in 3D nematics.
However, this is not so trivial from the viewpoint of topology, because $+1/2$ and $-1/2$ disclinations are topologically equivalent (homeomorphic) in 3D nematics \cite{Nakahara-Book2003,Chaikin.Lubensky-Book2000,deGennes.Prost-Book1995}.
Besides, unlike point disclinations, line disclinations have shapes and are deformable, giving additional potential sources of asymmetry.

\begin{figure}[t]
\includegraphics[width=\hsize,clip]{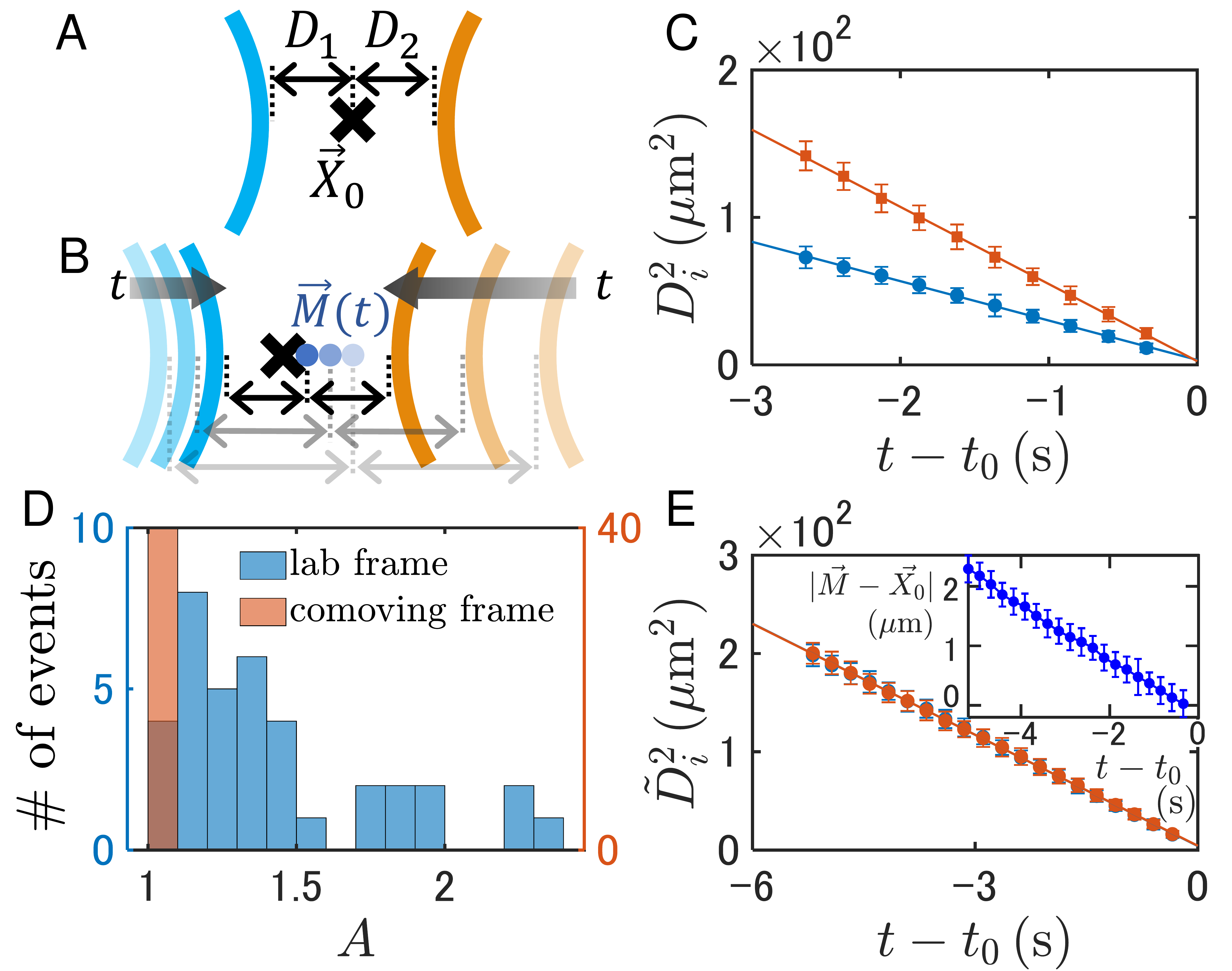}
\centering
\caption{Apparent asymmetry of in-plane reconnections. (\textit{A}) Definition of the distances $D_1(t)$ and $D_2(t)$ from the reconnection point $\vec{X}_0$. (\textit{B}) Sketch of the midpoint $\vec{M}(t)$ of the points on the disclinations closest to the reconnection point $\vec{X}_0$. (\textit{C}) Distance $D_i(t)$ measured in the laboratory frame, for an example pair of reconnecting disclinations. (\textit{D}) Histograms of the asymmetry parameter $A$ (the square root of the ratio of the two slopes in (\textit{C}), see text) measured in the laboratory frame (blue) and the co-moving frame (red). Note that three outliers are not displayed in the blue histogram, taking $A \approx 3.2$, $A \approx 5.5$ and $A \approx 40$ in the laboratory frame, but in the co-moving frame all data including those outliers fell in the first bin ($\max A = 1.03 \pm 0.01$). (\textit{E}) Distance $\tilde{D}_i(t)$ measured in the co-moving frame, for the pair shown in (\textit{C}). The inset shows the distance between the midpoint $\vec{M}(t)$ and the reconnection point $\vec{X}_0$ seen in the laboratory frame. The error bars in (\textit{C,E}) indicate the uncertainty evaluated from the Gaussian fitting and the search for the reconnection point used here (see Methods).}
\label{fig3}
\end{figure}

Here we inspected this asymmetry experimentally.
Instead of the distance $\delta(t)$ between reconnecting disclinations, we measured the distance between each disclination line and the reconnection point, $D_1(t)$ and $D_2(t)$ (\figref{fig3}A).
Plotting $D_i(t)^2$ against $t-t_0$, we found a power law $D_i(t) \simeq C_i|t-t_0|^{1/2}$ analogous to \eqref{eq:delta}, with coefficients $C_i$ that are typically asymmetric between the two disclinations (see \figref{fig3}C for an example).
The asymmetry was also clear from the defect motion (\supfigref{S-figS:defect3D}A).
Using the coefficients $C_i$, we define the asymmetry parameter $A$ by
\begin{equation}
    A \equiv \frac{\max\{C_1, C_2\}}{\min\{C_1, C_2\}} \label{eq:asymmetry}
\end{equation}
and determined it for each reconnection event.
By definition, $A=1$ for symmetric reconnections, and $A>1$ for asymmetric ones.
The histogram of $A$ (blue bars in \figref{fig3}D) shows that most in-plane reconnections appear to be significantly asymmetric.
We suspected that different curvature of the two disclination lines may contribute to this asymmetry, but this effect turned out to be minuscule (\supsecref{S-sec:curvature}).

\subsection*{Disappearance of asymmetry in the co-moving frame}
Let us now recall the fact that disclinations have extended line structures and also that the studied pairs were not the only defects present in the system.
It is therefore reasonable to consider that the reconnection dynamics may be affected by such extrinsic factors, which may induce flow and director changes superimposed to the intrinsic reconnection dynamics.
These effects are expected to add a drift to the intrinsic motion of reconnecting disclinations.
To evaluate this drift, we located the point on each disclination that was closest to the reconnection point, and inspected the motion of the midpoint $\vec{M}(t)$ of the pair of the closest points (\figref{fig3}B).
If the dynamics of the two disclinations are perfectly symmetric, the motion of this midpoint is the drift itself and will not show any singularity near the reconnection time.
If the dynamics is not symmetric, this midpoint will partly include the reconnection dynamics, showing the same singularity as $D_i(t) \propto |t-t_0|^{1/2}$.
This was indeed confirmed for the case of pair annihilation of 2D point disclinations reported by T\'oth \myetal \cite{Toth.etal-PRL2002} (\supfigref{S-figS:Toth}).
For 3D disclination lines, the behavior of $\vec{M}(t)$ is shown in the inset of \figref{fig3}E, for the pair displayed in \figref{fig3}C.
This clearly shows linear dependence on time, suggesting that the intrinsic dynamics of reconnection is actually symmetric.
Moreover, using the drift velocity $\vec{V}$ evaluated by fitting $\diff{\vec{M}}{t}$, we define the co-moving frame and measure the distance $\tilde{D}_i(t)$ between the closest point and the reconnection point in this co-moving frame.
The result shows that, remarkably, the reconnection dynamics in this co-moving frame is nearly perfectly symmetric (\figref{fig3}E and \supfigref{S-figS:defect3D}B).
We carried out this analysis for all reconnection events and for all cases the asymmetry parameter became very close to $1$ (red bar in \figref{fig3}D), the largest deviation being $A=1.03 \pm 0.01$.
Note that, though the asymmetry parameter $A$ is expected to be independent of the choice of reference frame in the limit $t \to t_0$, it does depend in practice, because the limit $t \to t_0$ is unreachable due to the finite time resolution of the observation.
Direct comparison of $D_i(t)$ and $\tilde{D}_i(t)$ (\supfigref{S-figS:CompDist}; also compare \figref{fig3}C and E) shows that the scaling law $D_i(t), \tilde{D}_i(t) \propto |t-t_0|^{1/2}$ appears longer in the co-moving frame than in the laboratory frame, indicating that the results in the co-moving frame are more reliable.
This will be supported in the next section, on the basis of the director configuration around the disclination pair.

\begin{figure*}[t]
\includegraphics[width=0.8\hsize,clip]{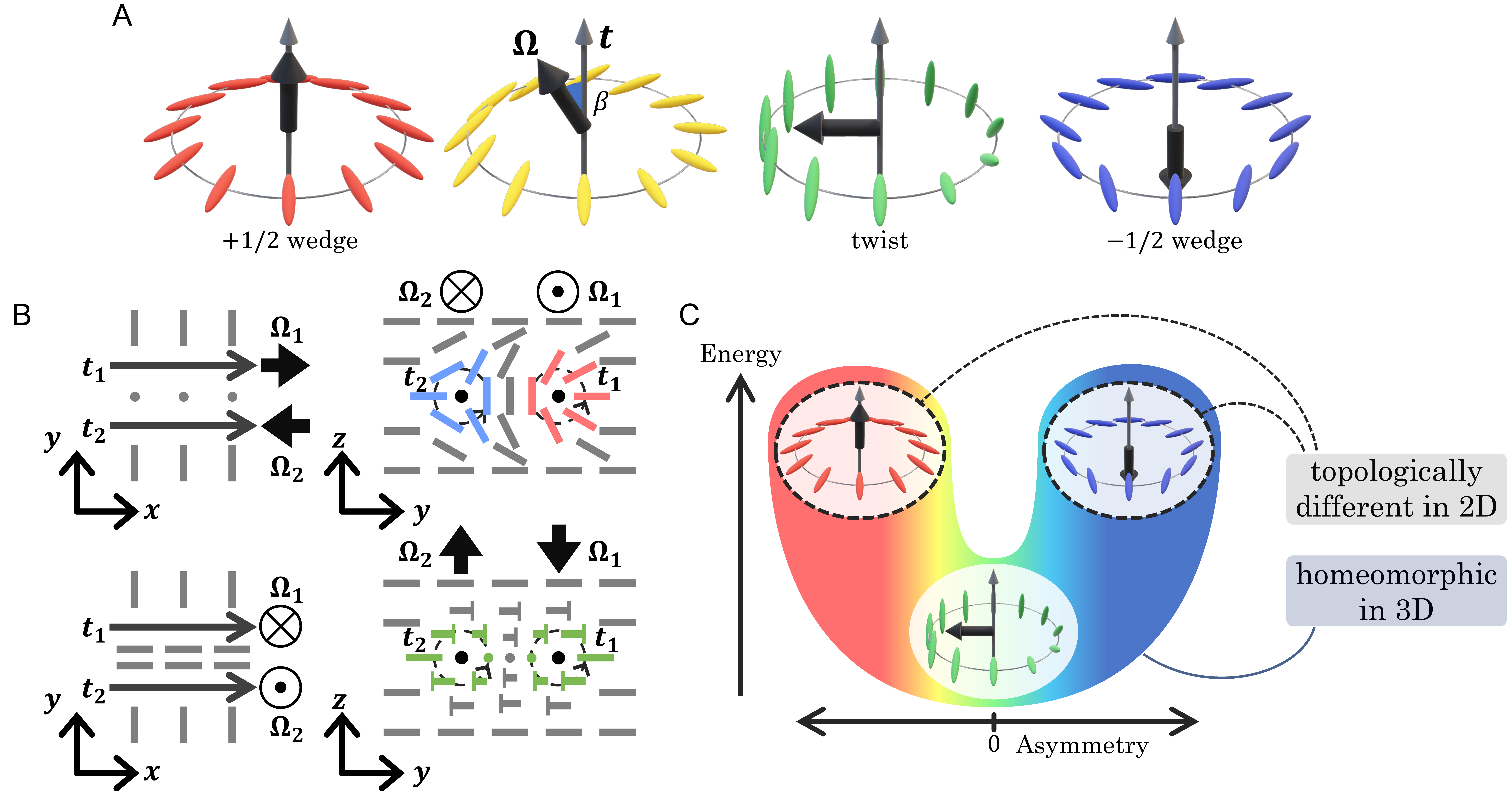}
\centering
\caption{Director configurations and asymmetry. (\textit{A}) Director field around a disclination for different $\vec{\Omega}$ (or $\beta$). (\textit{B}) Sketch of the director field around a wedge (top) and twist (bottom) disclination pair. The $xy$ (left) and $yz$ (right) cross sections are shown. The ``T'' symbols indicate that the directors are not in the plane of the cross section, with the T head above the paper. Note that the analyzed disclinations were not limited to those aligned in the x-direction as sketched here. (\textit{C}) Spontaneous symmetry restoring.}
\label{fig4}
\end{figure*}

\subsection*{Spontaneous symmetry restoring}
We have found that the asymmetry present in the 2D pair annihilation of $\pm 1/2$ point disclinations disappears for the in-plane reconnections of 3D disclination lines.
Obviously, if two disclination lines were straight and had $\pm 1/2$ director configurations around (\figref{fig4}A left and right), this pair would exhibit the same asymmetry as its 2D counterpart.
However, since $+1/2$ and $-1/2$ disclinations are homeomorphic in 3D \cite{Nakahara-Book2003,Chaikin.Lubensky-Book2000,deGennes.Prost-Book1995}, the director can actually take an intermediate configuration that continuously transforms between these two limiting structures (\figref{fig4}A).
More precisely, the winding of the director around a disclination line is \textit{not} characterized by the winding number, but by a unit vector that specifies the rotation axis of the director, denoted by $\vec{\Omega}$ (see, e.g., \cite{Long.etal-SM2021,Schimming.Vinals-SM2022}).
With the unit tangent vector $\vec{t}$ whose head and tail are set arbitrarily, the director rotates right-handed by $180^\circ$ in the plane perpendicular to $\vec{\Omega}$, along a closed path that turns right-handed about the tangent vector $\vec{t}$.
If $\vec{\Omega}=\vec{t}$ (or the angle $\beta \equiv \cos^{-1}(\vec{\Omega}\cdot\vec{t})=0$), the director is essentially in the plane perpendicular to the disclination line and the defect is equivalent to a $+1/2$ point disclination in that plane.
Similarly, if $\vec{\Omega}=-\vec{t}$ ($\beta=\pi$), it is equivalent to a $-1/2$ point disclination.
These two limiting structures, called the wedge disclinations, are interpolated continuously by intermediate $\beta$.
In particular, if $\vec{\Omega}\perp\vec{t}$ ($\beta=\pi/2$), the director purely twists around the defect; hence, it is called a twist disclination.

Now, for a pair of reconnecting disclinations in plane, we have two tangent vectors $\vec{t}_1$ and $\vec{t}_2$ which are parallel near the reconnection point, so that we choose $\vec{t}_1=\vec{t}_2$.
Then it is reasonable to assume $\vec{\Omega}_1 = -\vec{\Omega}_2$ ($\beta_2=\pi-\beta_1$) so that the disclinations may attract each other most effectively, as inferred from the disclination dynamics derived in the theoretical studies \cite{Long.etal-SM2021,Schimming.Vinals-SM2022}. Indeed, $\vec{\Omega}_1 =-\vec{\Omega}_2$ is expected in order to reduce the elastic energy cost due to the existence of the disclination pair.
This leaves one free parameter, $\beta_1$ (or $\vec{\Omega}_1)$.
If $\beta_1=0$ or $\pi$, we have a pair of $\pm 1/2$ wedge disclinations (\figref{fig4}B top), which is equivalent to a pair of annihilating point disclinations in 2D nematics and therefore approach asymmetrically \cite{Toth.etal-PRL2002,Svensek.Zumer-PRE2002}.
By contrast, if $\beta=\pi/2$, we have a pair of twist disclinations (\figref{fig4}B bottom) with $\pi$ rotationally symmetric director field; in this case, the dynamics of the two disclinations should also be symmetric.

Our experimental results of the vanishing asymmetry suggest that all disclination pairs we observed spontaneously took the symmetric twist configurations.
This can be attributed to the anisotropic elasticity of liquid crystal: bulk deformation of the director can be decomposed into splay, twist, and bend deformations, characterized by different elastic constants denoted by $K_1$, $K_2$, and $K_3$, respectively \cite{deGennes.Prost-Book1995}.
For the mesogen used here, MLC-2037, these are $K_1=11.6\unit{pN}, K_2=6.1\pm0.5\unit{pN}, K_3=13.2\unit{pN}$ (\suptblref{S-tbl:MLC2037}, and Methods).
Similarly to other typical mesogens, the elastic constant for twist deformations is lower than that for splay and bend deformations.
Then it follows that the twist configuration of the disclination pair (\figref{fig4}B bottom) is energetically favored over the wedge configuration (\figref{fig4}B top), which involves splay and bend deformations of the director field.
This explains why the twist configuration seemed to be exclusively observed in our experiments, accounting for the disappearance of the asymmetry.

Moreover, we confirm the realization of the twist configuration via the coefficient $C$ of the power law in \eqref{eq:delta}, as follows.
Balancing the drag force $J\gamma_1(\dot{\delta}/2)$ according to Geurst \myetal \cite{Geurst.etal-JPP1975}, with a dimensionless coefficient $J \approx 1.9$ and the rotational viscosity $\gamma_1$, and the attractive force $\pi K/2\delta$ exerted to the pair with $\vec{\Omega}_1 = -\vec{\Omega}_2$ under the one-constant approximation $K_1=K_2=K_3 \equiv K$ \cite{Long.etal-SM2021}, we obtain
\begin{equation}
    C^2 = \frac{2\pi K}{J \gamma_1}.  \label{eq:CK}
\end{equation}
However, since the actual elastic constant is anisotropic, \eqref{eq:CK} is expected to hold with $K \approx K_1, K_3$ for the wedge configuration and $K \approx K_2$ for the twist one. 
From our data (\figref{fig2}) and $\gamma_1 = 132\unit{mPa\cdot s}$ for our mesogen (\suptblref{S-tbl:MLC2037}), we obtain $C^2 = 151 \pm 27\unit{\mu m^2/s}$ (the range of error being the standard deviation), which is close to the value for the twist configuration, $C^2 \approx 1.5 \times 10^2\unit{\mu m^2/s}$, instead of that for the wedge one, $C^2 \approx 3.1 \times 10^2\unit{\mu m^2/s}$.
This supports the realization of the twist configuration in the disclination pairs we observed, as well as the resulting, vanishing asymmetry we found in the co-moving frame.

We also demonstrate the realization of the twist configuration in a more direct but destructive manner, through the pattern of the electroconvection induced in the sample.
It is known that nematic liquid crystal with negative dielectric anisotropy and positive conductivity anisotropy, such as the one used in the present work, shows roll convection under a moderate applied voltage \cite{deGennes.Prost-Book1995}.
The direction of the rolls is determined by the director near the midplane; it is normal to the director if the director is parallel to the cell surfaces, while patches of rolls of different directions are formed if the director is perpendicular to the surfaces.
Our observation reveals that the region between disclinations shows convection rolls normal to those in the outer region (\supfigref{S-figS:WD}), indicating the twist director configuration as sketched in \figref{fig4}B bottom (see \supsecref{S-sec:WD} for details).

While the twist configuration is expected from the energy viewpoint, it is important to note that such a lowest energy configuration is to describe the equilibrium state, while our observations deal with relaxation to it.
Upon quenching from the turbulent state, we expect that there exist various types of disclinations, from wedge to twist and in between.
However, since all these configurations are homeomorphic, disclinations are allowed to change the configurations continuously, toward the lowest energy state, i.e., the twist configuration.
This is not possible for 2D nematics, for which $+1/2$ and $-1/2$ disclinations are topologically distinct.
Mathematically, this is a consequence of the different homotopy groups between 2D and 3D nematics \cite{Nakahara-Book2003,deGennes.Prost-Book1995}.
For 2D, it is $\pi_1(\mathbb{R}P^1) = \mathbb{Z}$, which distinguishes all different winding numbers.
In contrast, its 3D counterpart is $\pi_1(\mathbb{R}P^2) = \mathbb{Z}_2$, which now distinguishes only the absence and the presence of a nontrivial defect configuration. 
In particular, wedge disclinations of winding number $\pm 1/2$ are now identified through continuous transformation, with the symmetric twist state found in the middle, at the lowest energy (\figref{fig4}C).
This results in the realization of the symmetric reconnection dynamics, as we observed experimentally.

At this point, it is not difficult to generalize the argument.
If the space of the order parameter field accommodating topological defects of interest is related to real space, such as the nematic case ($\mathbb{R}P^{d-1}$ for $d$-dimensional space), the corresponding homotopy group may also depend on the dimensionality.
In the case where there exist two asymmetric structures that are topologically distinct in a lower dimension but become homeomorphic in a higher dimension, such as the case of $\pm 1/2$ nematic disclinations, the defect in the higher dimension is allowed to take an intermediate structure that continuously interpolates the two asymmetric analogues of those in the lower dimension.
Then it is likely that a symmetric intermediate structure exists. If it is energetically favored, the asymmetry present in the lower dimension will tend to disappear in the higher dimension spontaneously.
In brief, if two topologically distinguished and asymmetric configurations in a lower dimension become homeomorphic in a higher dimension, and if the newly allowed symmetric configuration is energetically favorable, symmetry is spontaneously restored. 
The symmetry in the structure results in the symmetry in the dynamics.
Our results on reconnecting nematic disclinations constitute a clear example of such spontaneous restoring of symmetry.

\subsection*{Concluding remarks}
We carried out a direct confocal observation of disclination dynamics in 3D nematics, using the accumulation of fluorescent dyes to disclinations.
Our method successfully resolved characteristic dynamics of disclination lines, such as reconnections and loop shrinkage.
Studying in-plane reconnection events in depth, we demonstrated the distance-time scaling law (\eqref{eq:delta}) predicted for straight disclination pairs, despite the curved shape of the observed disclinations.
Moreover, we revealed that the dynamics of reconnecting disclinations is only deceptively asymmetric in the 3D case, being actually symmetric in the co-moving frame.
This is explained by the spontaneous realization of symmetric twist configurations, which is energetically favored because of the lower twist elasticity.
These observations led us to propose a mechanism of such spontaneous symmetry restoring, from a general viewpoint of topology and energy.
In this regard, it is important to investigate the generality and limitation of this concept in future studies.
The first step would be to study intersecting reconnections.
Although we restricted the analysis to in-plane reconnections in the present work, our argument suggests that the spontaneous symmetry restoring also holds for intersecting reconnections. Simulations in Ref. \cite{Schimming.Vinals-SM2022} showed that at least the same scaling law holds for intersecting reconnections.
Furthermore, it is of prominent importance to test the fate of the symmetry restoring in the case where the condition $K_2 < K_1, K_3$ is not satisfied and consequently the twist configuration does not correspond to the lowest energy state. Such a situation may be realized by using the divergence of $K_2$ near the nematic-smectic transition \cite{deGennes.Prost-Book1995}, by the drop of $K_3$ near the transition to the twist-bend nematic phase \cite{Adlm.etal-PRL2013}, or by using nematic discotic liquid crystals \cite{Osipov.etal-MP-1993}.

Since the concept of topological defects is universal, it is important to think of similarities and dissimilarities in defect properties across different disciplines of physics.
For example, quantum vortices in superfluid $^4$He are known to have similar interaction energy and show the same scaling law of $\delta(t)$ (\eqref{eq:delta}) as observed experimentally \cite{Bewley.etal-PNAS2008,Fonda.etal-PNAS2014,Minowa.etal-SA2022}, while the corresponding homotogy group is different and the rotation axis $\vec{\Omega}$, if defined analogously, is fixed to $\vec{\Omega} = \vec{t}$ or $-\vec{t}$.
Further, it is tempting to seek for examples of spontaneous symmetry restoring we proposed in this work.
Disgyration of superfluid $^3$He is particularly interesting in this context, for which the homotopy group is $\pi_1(SO(2))=\mathbb{Z}$ for 2D and $\pi_1(SO(3))=\mathbb{Z}_2$ for 3D, and its asymmetric structures as well as energy have been thoroughly discussed \cite{Vollhardt.Woelfle-book2013}.
We hope that such approaches to general mechanisms will accelerate multidisciplinary understanding of topological defects and that the visualization of nematic disclination dynamics reported here will be a useful tool in this line.

\matmethods{

\subsection*{Sample preparation and defect generation}  \label{sec:expcond}

The experimental sample was prepared as follows for the main results on the in-plane reconnections, while changes for other experiments are described in the end of this subsection. 
The liquid crystal sample was nematic compound MLC-2037 (Merck, discontinued product), doped with $0.5\unit{wt\%}$ of tetra-$n$-butylammonium bromide and $0.005\unit{wt\%}$ of fluorescent dye Coumarin 545T.
The mesogen MLC-2037 was chosen for its low birefringence $\Delta n=0.0649$ and negative dielectric anisotropy $\Delta\epsilon<0$ (\suptblref{S-tbl:MLC2037}), the latter of which was used to induce electroconvection to generate disclinations \cite{deGennes.Prost-Book1995}.
The sample was introduced to a hand-made cell, which consists of a coverslip and a glass plate both coated with indium tin oxide, and $130\unit{\mu{}m}$ thick polyimide tapes used as spacers.
The inner surfaces were coated with polyvinyl alcohol and rubbed to realize a homogeneous planar alignment.

To study disclination dynamics, we generated a large density of disclinations by applying an alternating electric field (root-mean-square amplitude $150\unit{V}$, frequency $50\unit{Hz}$) to the sample, inducing an electrohydrodynamic turbulence called the dynamic scattering mode 2 \cite{deGennes.Prost-Book1995,Kai.Zimmermann-PTPS1989}.
Then we removed the electric field and observed relaxation of disclinations by a confocal laser scanning microscope (Leica SP8, objective 20x, NA 0.75, oil immersion), equipped with a resonant scanner working at $8\unit{kHz}$ and a piezo objective scanner.
The fluorescent dyes were excited at $488\unit{nm}$ by laser light polarized in the direction perpendicular to the nematic easy axis, represented by $x$ and $y$-axes, respectively.
The fluorescence signal in the range between $500$ and $600\unit{nm}$ was confocally detected by a photomultiplier tube detector (pinhole size $20\unit{\mu{}m}$, roughly $0.35$ times the Airy unit).
The voxel size in the $xy$ plane was $0.91\unit{\mu{}m}$ and the spacing between $z$ slices was $1\unit{\mu{}m}$.
The number of voxels was $512, 128, 21$ in the $x, y, z$ directions, respectively.
The time interval between consecutive confocal images was $0.255\unit{s}$.
Compared with this, the time needed for fluorescent dyes to follow the evolution of disclination lines is expected to be much shorter, which we evaluate to be roughly $1\unit{ms}$, using a length scale $0.33\unit{\mu{}m}$ reported as the apparent size of dye accumulation in \cite{Ohzono.etal-SR2016} and the typical value of the diffusion coefficient of dye molecules, $10^{-10}\unit{m^2/s}$ \cite{Lavrentovich-PJP2003,Blinov.Chigrinov-Book1994}.

Below we describe experimental conditions used for other observations.
Conditions and parameters that are not specified below were kept unchanged from those for the in-plane reconnections.
The intersecting reconnection displayed in \figref{fig1}E was observed in a sample that contained $0.25\unit{wt\%}$ of tetra-$n$-butylammonium bromide and $0.2\unit{wt\%}$ of Coumarin 545T. 
After an alternating voltage of root-mean-square amplitude $150\unit{V}$ and frequency $100\unit{Hz}$ was removed, we observed the intersecting reconnection in a manner similar to the case of the in-plane reconnections, except that the number of voxels was $512 \times 64 \times 36$ and the time interval was $0.300\unit{s}$.
The loop shrinkage displayed in \figref{fig1}F was observed in a sample that contained $0.05\unit{wt\%}$ of tetra-$n$-butylammonium bromide and $0.005\unit{wt\%}$ of Coumarin 545T.

\subsection*{Image analysis} \label{sec:image}

The data acquired by the confocal microscope were the fluorescence intensity detected at 3D position $(x,y,z)$ and time $t$.
Using the 3D image at each time, we obtained cross sections and extracted the coordinates of the disclinations as follows.
First we chose the cross sections to use, either in the $xz$ plane or in the $yz$ plane, chosen so that the cross sections become closer to perpendicular to the disclination lines.
In each cross section, the two disclinations appear as bright spots.
These bright spots were fitted by a Gaussian function to obtain the coordinates of the spot centers.
Repeating this over all cross sections, we obtained a set of 3D coordinates along each disclination line.
The closest distance $\delta(t)$ between two disclinations (\figref{fig2}) was directly determined from these coordinates.

The time and the position of each reconnection event, as well as the distance $D_i(t)$ $(i=1,2)$ of a disclination from the reconnection point, were determined as follows.
For the reconnection time $t_0$, we determined it from the 2D image constructed from the transmitted excitation laser, to benefit from the finer time resolution than that of the confocal images.
For the position $\vec{X}_0 = (X_0,Y_0,Z_0)$ of the reconnection point, we first approximately located it from the series of transmitted and confocal images ($X_0$ and $Y_0$ from the transmitted images, $Z_0$ from the confocal images).
Using this and the coordinates of the disclinations, we could evaluate the distance $D_i(t)$ in the laboratory frame, but for the analysis presented in the paper, we evaluated $D_i(t)$ more precisely in the following manner.
First, we fitted the 3D coordinates of disclinations by smoothing splines, to reduce the noise and to interpolate the lines appropriately.
In general, smoothing splines $s(x)$ for a data set $(x_i, y_i)$ are such a function that minimizes
\begin{equation}
  p\sum_i w_i \qty(y_i -s(x_i))^2 +(1-p)\int \qty(\dv[2]{s}{x})^2 \dd{x},
\end{equation}
with a smoothing parameter $p$ and a weight $w_i$, which is set to be $1$ here.
By adjusting $p$, we obtained smoothing splines that reproduced the defect shape without high wave number components, for the two coordinates that spanned the cross sections (i.e., for $xz$ cross sections, the obtained smoothing splines were $x(y)$ and $z(y)$).
Then, we also refined the estimate of the reconnection point $\vec{X}_0$, by using the coordinates of the disclinations before the moment of the reconnection.
Specifically, we determined $\vec{X}_0$ in such a way that the scaling $D_i(t) \simeq C_i|t-t_0|^{1/2}$ is satisfied most precisely in a time period before the reconnection, under the constraint that $\vec{X}_0$ is not changed by more than $3\unit{\mu{}m}$ from the first rough estimate.
This was done by evaluating $D_i(t)$ for each candidate position $\vec{X}_0$ in the range within $3\unit{\mu{}m}$, fitting it to $D_i(t)^2 = a_i|t-t_0| + b_i$, and choosing the candidate $\vec{X}_0$ that minimizes $b_1^2 + b_2^2$.
The distance $\tilde{D}_i(t)$ in the co-moving frame was also determined analogously, by using the position $\vec{X}_0$ that drifts with the velocity of the co-moving frame.

The errors in the estimates of $\delta(t), D_i(t), \tilde{D}_i(t)$ were evaluated as follows.
For $\delta(t)$, the error bars in \figref{fig2} indicate the square root of the sum of the squares of the uncertainties in all coordinates of the two closest points.
For the coordinates in the cross section, we used the $95\%$ confidence interval of the Gaussian fitting as the uncertainty; for the other coordinate, we used half the thickness of the cross section, i.e., the voxel size, as the uncertainty.
For $D_i(t)$ and $\tilde{D}_i(t)$ (\figref{fig3}C,E), the errors were evaluated from the uncertainties in the coordinates of the reconnection point and the closest point on the disclination line, again by the square root of the sum of the squares.
The uncertainties in the coordinates of the reconnection point were considered to be half the size of the scanned region described above.
For the uncertainties in the coordinates of the closest point on the disclination line, since the closest point was located on the smoothing spline, we only considered the uncertainties ($95\%$ confidence interval) for the coordinates in the cross section that is closest to the closest point on the spline.

\subsection*{Estimation of $K_2$} \label{sec:K2}

The twist elastic constant $K_2$ of MLC-2037 was evaluated by using the Fr\'eedericksz transition under an external magnetic field \cite{deGennes.Prost-Book1995}.
The Fr\'eedericksz transition point $H^\mathrm{c}_i$ corresponding to the elastic constant $K_i$ is given by 
\begin{equation}
    H^\mathrm{c}_i = \frac{\pi}{d}\sqrt{\frac{K_i}{\Delta \chi}},
\end{equation}
where $d$ is the cell thickness and $\Delta\chi$ is the magnetic anisotropy.
For MLC-2037, $\Delta\chi$ was unknown but $K_1$ is known (see \suptblref{S-tbl:MLC2037}). Therefore, we measured the Fr\'eedericksz transition for both the splay and twist configurations, obtaining $H^\mathrm{c}_1$ and $H^\mathrm{c}_2$, and used the ratio
\begin{equation}
    \frac{H^\mathrm{c}_2}{H^\mathrm{c}_1} = \sqrt{\frac{K_2}{K_1}}  \label{eq:K2}
\end{equation}
to determine $K_2$ from $K_1$.

We used a ready-made cell with homogeneous planar alignment (EHC, KSRO-25/B107M6NTS, $d = 25\unit{\mu{}m}$) filled with MLC-2037.
Using a superconducting magnet, we applied a magnetic field perpendicular to the cell surface for the splay configuration, and parallel to the cell surface but perpendicular to the easy axis for the twist configuration.
The Fr\'eedericksz transition point was determined by measuring the retardation change, through the transmitted light intensity that changed in a swept magnetic flux density $B$ under crossed Nicols (\supfigref{S-figS:K2}).
The measurement for the twist configuration was performed at oblique incidence ($5^{\circ}$) to reduce the effect of polarization rotation \cite{Chandrasekhar-Book1993,Karat-thesis1977}.

We measured the Fr\'eedericksz transition eight times for the splay configuration and three times for the twist configuration.
The light source was either a halogen lamp or a light-emitting diode.
For each measurement, we determined the transition point twice, when the magnetic field was increased and decreased.
As a result, we obtained a total of $16$ estimates of the transition point $B^\mathrm{c}_1$ and $6$ estimates of $B^\mathrm{c}_2$.
By using all of them, we determined our final estimates at $B^\mathrm{c}_1 = 4.4 \pm 0.1\unit{T}$ and $B^\mathrm{c}_2 = 3.2 \pm 0.1\unit{T}$.
Then it follows, by using \eqref{eq:K2} and $K_1 = 11.6\unit{pN}$ (\suptblref{S-tbl:MLC2037}), that $K_2 = 6.1 \pm 0.5\unit{pN}$.

\subsection*{Data availability}

Analysis results have been deposited in figshare (\url{https://doi.org/10.6084/m9.figshare.21130301}). 

%

}
\showmatmethods{} 

\acknow{
We are greatly indebted to F.~Araoka at RIKEN CEMS for providing the experimental setup for determination of $K_2$ and for his help to measure that of MLC-2037.
We are grateful to M.~Tsubota for his suggestion to study the coefficient $C$ of \eqref{eq:delta}, and to M.~Kobayashi for drawing the authors' attention to disgyration.
We thank G.~T\'oth, C.~Denniston, and J.~M.~Yeomans for allowing us to use and analyze their data in Ref.~\cite{Toth.etal-PRL2002}.
We acknowledge the material data of MLC-2037 in \suptblref{S-tbl:MLC2037} provided by Merck and their permission to present them in this work.
We also thank C.~Denniston, J.-i.~Fukuda, O.~Ishikawa, T.~Ohzono, K.~Katoh, J.~V.~Selinger, R.~L.~B.~Selinger, and H. Watanabe for useful discussions. 
This work is supported in part by JST PRESTO (Grant No. JPMJPR18L6), by KAKENHI from Japan Society for the Promotion of Science (Grant Nos. JP19H05800, JP19H05144, JP20H01826, JP22J12144), by JSR Fellowship (The University of Tokyo), and by FoPM, WINGS Program (The University of Tokyo.).
}
\showacknow{} 

\bibliography{ref}

\end{document}


\maketitle

\SItext

\subsection*{Evaluation of the curvature effect}  \label{sec:curvature}

Here we evaluate the effect of different curvature of the two disclinations, as a potential source of the apparently asymmetric dynamics observed in the laboratory frame.
It is well known that disclinations have a line tension $T$ apart from a logarithmic correction, which is roughly the elastic constant $K$ under the one-constant approximation, $T \approx K$ \cite{deGennes.Prost-Book1995,Geurst.etal-JPP1975}.
This implies curvature-driven force per length, $T/R$, exerted to a curved disclination, where $R$ is the local radius of curvature.
Geurst \myetal \cite{Geurst.etal-JPP1975} described this phenomenologically and proposed that the contribution of this force to the disclination velocity is approximately $v^\mathrm{curv} \approx K/\gamma_1 R$, where $\gamma_1$ is the rotational viscosity.
We used this to compensate for the effect of curvature.
The radius of curvature at each point on the disclination line was determined from the smoothing splines we obtained in the image analysis (Methods).
First, we used the power law $D_i(t) \simeq C_i |t-t_0|^{1/2}$ in the laboratory frame and evaluated the velocity $\diff{D_i}{t}$.
The solid lines in \supfigref{figS:curv}B display $\left| \diff{D_i}{t} \right|$ for the pair shown in \supfigref{figS:curv}A (same as \figref{M-fig3}C in the main paper).
We then evaluated the curvature contribution $v^\mathrm{curv}_i = K/\gamma_1 R_i$, with $K=K_3$ for evaluating the largest possible value, using the radius of curvature $R_i$ measured at the point closest to the reconnection point.
\supfigref{figS:curv}B shows the obtained values of $\left| \diff{D_i}{t} \right| + v^\mathrm{curv}_i$ by the open symbols,
as well as the results of fitting by
\begin{equation}
    \left|\diff{D_i}{t}\right| + v^\mathrm{curv}_i \simeq \frac{C_i}{2}|t-t_0|^{-1/2} \label{eq:speed_curv}
\end{equation}
by the dashed lines.
Comparing the speed with and without compensation of the curvature effect (dashed and solid lines in \supfigref{figS:curv}B, respectively), we find that the contribution of the curvature effect is much smaller than the observed asymmetry.
We confirmed this for the ensemble of the reconnection events, by re-evaluating the asymmetry parameter $A$ [\eqref{M-eq:asymmetry} in the main paper], i.e., 
\begin{equation}
    A \equiv \frac{\max\{C_1, C_2\}}{\min\{C_1, C_2\}},
\end{equation}
 and making a histogram (red bars in \supfigref{figS:curv}C).
Therefore, we conclude that the asymmetry observed in the laboratory frame is far stronger than the asymmetry induced by curvature.

\subsection*{Observation of roll convection during the relaxation process}  \label{sec:WD}

We demonstrate that the director field is twisted in the region between disclinations, by using the pattern of the electroconvection induced in the sample.
It is known that nematic liquid crystal with negative dielectric anisotropy and positive conductivity anisotropy, such as the one used in the present work, shows roll convection under a moderate applied voltage \cite{deGennes.Prost-Book1995}.
The roll convection results in the appearance of an evenly spaced stripe pattern in the transmitted light microscopy, called the Williams domain \cite{deGennes.Prost-Book1995}.
The direction of the stripe is determined by the director near the midplane between the top and bottom surfaces.
If the director is parallel to the cell surfaces, the stripe is normal to the director.
By contrast, if the director is perpendicular to the surfaces, a patchy pattern of regions with different stripe directions, which evolve chaotically in space and time, is known to appear \cite{Hidaka.etal-JPSJ1997}.
Therefore, observation of the electroconvection pattern may give a direct hint on the direction of the director, which will tell us whether the configuration around disclination pairs is closer to the wedge configuration or the twist configuration (\figref{M-fig4}B top or bottom, respectively).

The observation was carried out as follows. Liquid crystal sample MLC-2037 was doped with $0.5\unit{wt\%}$ of tetra-$n$-butylammonium bromide, filtered, and sealed in the $130\unit{\mu m}$ thick sandwich cell.
As in the main experiments, we applied an alternating voltage (here at root-mean-square (RMS) amplitude $149\unit{V}$ and frequency $40\unit{Hz}$) to the sample to generate a large density of disclinations in the turbulent state and then switched off the voltage.
The disclinations showed a relaxation process including reconnections and loop shrinkage as in the main experiments.
Then, at a given moment, we applied a voltage of $14\unit{V}$ RMS to induce roll convection.
The applied voltage was chosen so that it is sufficiently higher than the onset voltage of the roll convection in our sample, $8\unit{V}$ at the frequency we used, to generate the stripe pattern fast enough. 
We did this observation by transmitted light microscopy, using an inverted microscope (Olympus IX73, Objective 10x NA 0.30, dry).

A typical observation result is shown in \figref{figS:WD}.
Figure\,\ref{figS:WD}A shows a pair of disclinations, which are about to reconnect and which we focus on.
We then applied $14\unit{V}$ RMS and observed the stripe pattern shown in \figref{figS:WD}B.
The observed stripe pattern between the disclination pair is not patchy but aligned uniformly and perpendicularly to the stripes outside the pair, as sketched in \figref{figS:WD}C.
This demonstrates that the director field between the disclinations took the twist configuration. 
Note that, near the reconnection point, the two disclinations were too close to observe the stripe pattern in between, but it is safe to assume that the director there was not significantly different from the other region where the stripes were clearly visible between the disclination pair, because no singular nor nonsingular defects were present nearby.
In summary, this observation demonstrates that typical reconnecting disclination pairs indeed took the twist director configuration.


\begin{table}
 \centering
 \caption{Material parameters of MLC-2037 at temperature $20\unit{^\circ{}C}$, provided by Merck. Note that MLC-2037 is a discontinued product.}
 \label{tbl:MLC2037}
 \catcode`?=\active \def?{\phantom{0}}
 \begin{tabular}{lll} \hline\hline
  refractive index & (extraordinary) & $n_\mathrm{e}=1.5371$ \\
  & (ordinary) & $n_\mathrm{o}= 1.4722$ \\
  & (anisotropy) & $\Delta n \equiv n_\mathrm{e}-n_\mathrm{o} = 0.0649$ \\ \hline
  dielectric constant & (parallel) & $\epsilon_\parallel = 3.6$ \\
  & (perpendicular) & $\epsilon_\perp=6.7$ \\
  & (anisotropy) & $\Delta\epsilon \equiv \epsilon_\parallel-\epsilon_\perp=-3.1$ \\ \hline
  elastic constants & (splay) & $K_1=11.6\unit{pN}$ \\
  & (bend) & $K_3=13.2\unit{pN}$ \\ \hline
  rotational viscosity & & $\gamma_1=132\unit{mPa \cdot s}$ \\ \hline\hline
 \end{tabular}
\end{table}


\begin{figure}
\centering
\includegraphics[width=0.5\hsize,clip]{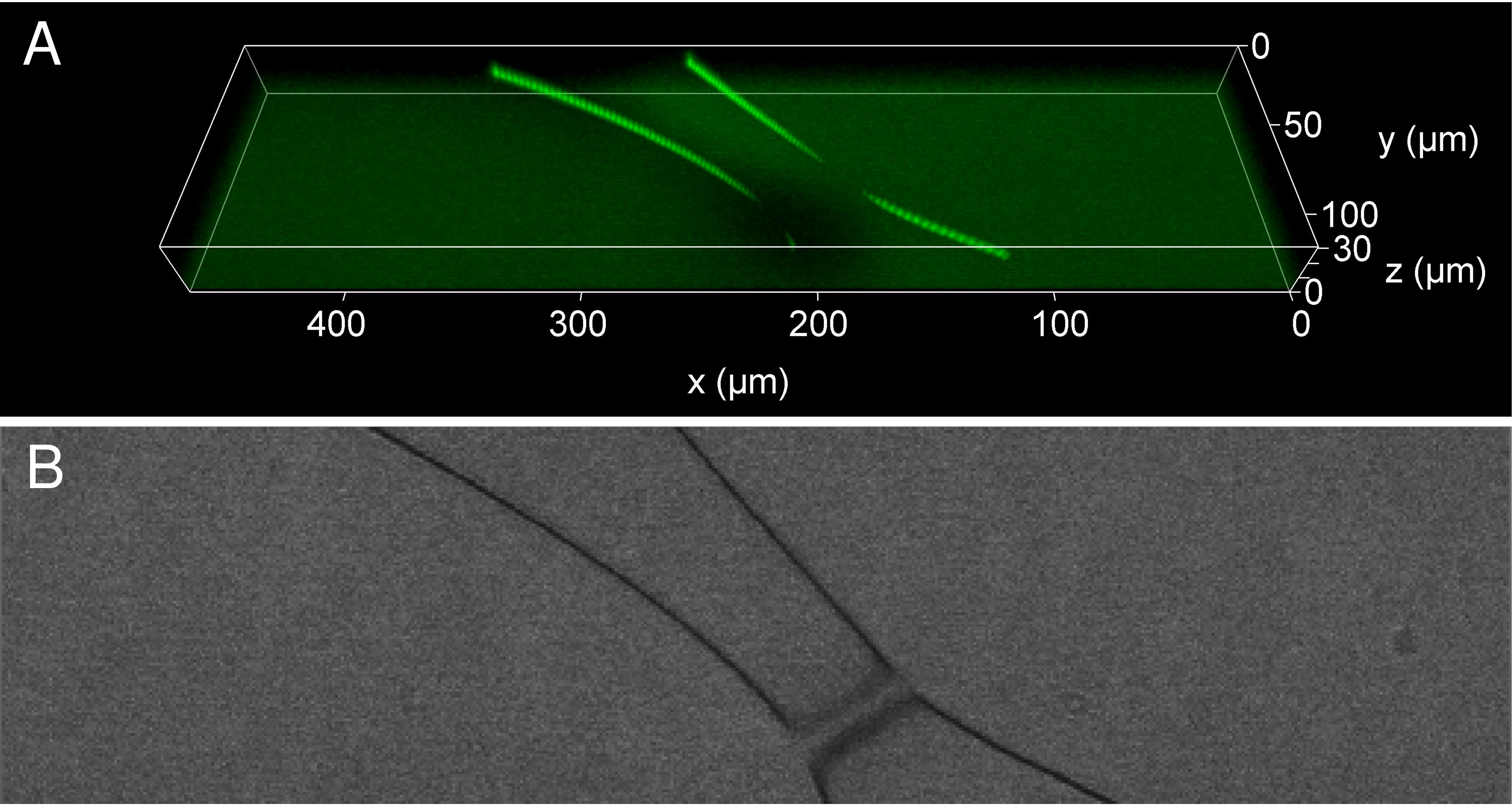}
\caption{Nonsingular disclination bridging a pair of singular disclinations. Confocal (\textit{A}) and bright-field (\textit{B}) images taken at the same moment are shown. The number of voxels was $512, 128, 31$ in the $x,y,z$ directions, respectively. The applied voltage before removal was of $150\unit{V}$ and $50\unit{Hz}$. Other experimental conditions were the same as those used for the intersecting reconnection in \figref{M-fig1}E, except that the time interval of the confocal images was $0.374\unit{s}$ here.}
\label{figS1}
\end{figure}


\begin{figure}
\includegraphics[width=0.8\hsize,clip]{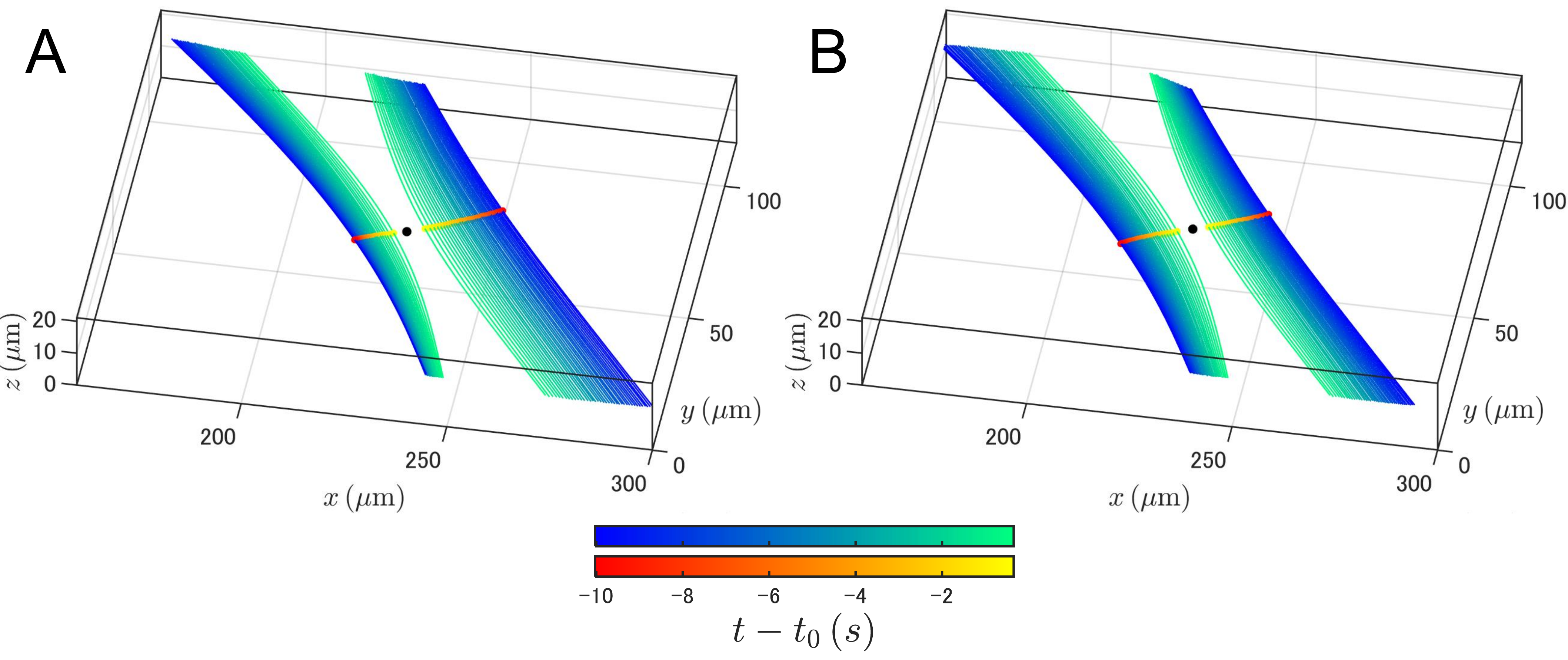}
\centering
\caption{
Motion of a disclination pair in the laboratory frame (\textit{A}) and in the co-moving frame (\textit{B}). The motion of disclination lines is indicated by cold colors and that of the points closest to the reconnection point (black dot) by warm colors. The color gradients indicate the time. The disclination pair used in \figref{M-fig3}C,E is shown. 
}
\label{figS:defect3D}
\end{figure}


\begin{figure}
\includegraphics[width=\hsize,clip]{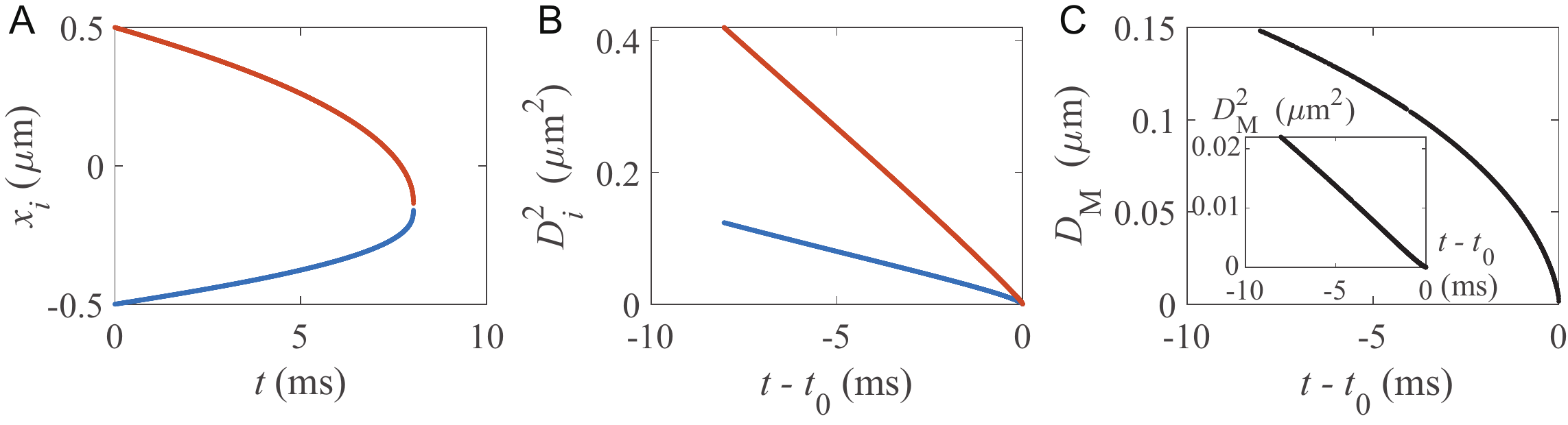}
\centering
\caption{
Analysis of the case of point disclinations in 2D nematics. The data in Fig.2(a) of Ref.~\cite{Toth.etal-PRL2002} were extracted and analyzed here, with permission of the authors. (\textit{A}) Time evolution of the positions $x_i$ of a $+1/2$ (red) and a $-1/2$ (blue) point disclination before the pair annihilation. (\textit{B}) The squared distance $D_i(t)^2$ of each defect from the reconnection point. The asymmetry parameter is estimated at $A=2.0 \pm 0.4$. (\textit{C}) The distance $D_M$ between the midpoint $(x_1+x_2)/2$ and the reconnection point. The inset shows $D_M^2$.}
\label{figS:Toth}
\end{figure}


\begin{figure}
\includegraphics[width=0.7\hsize,clip]{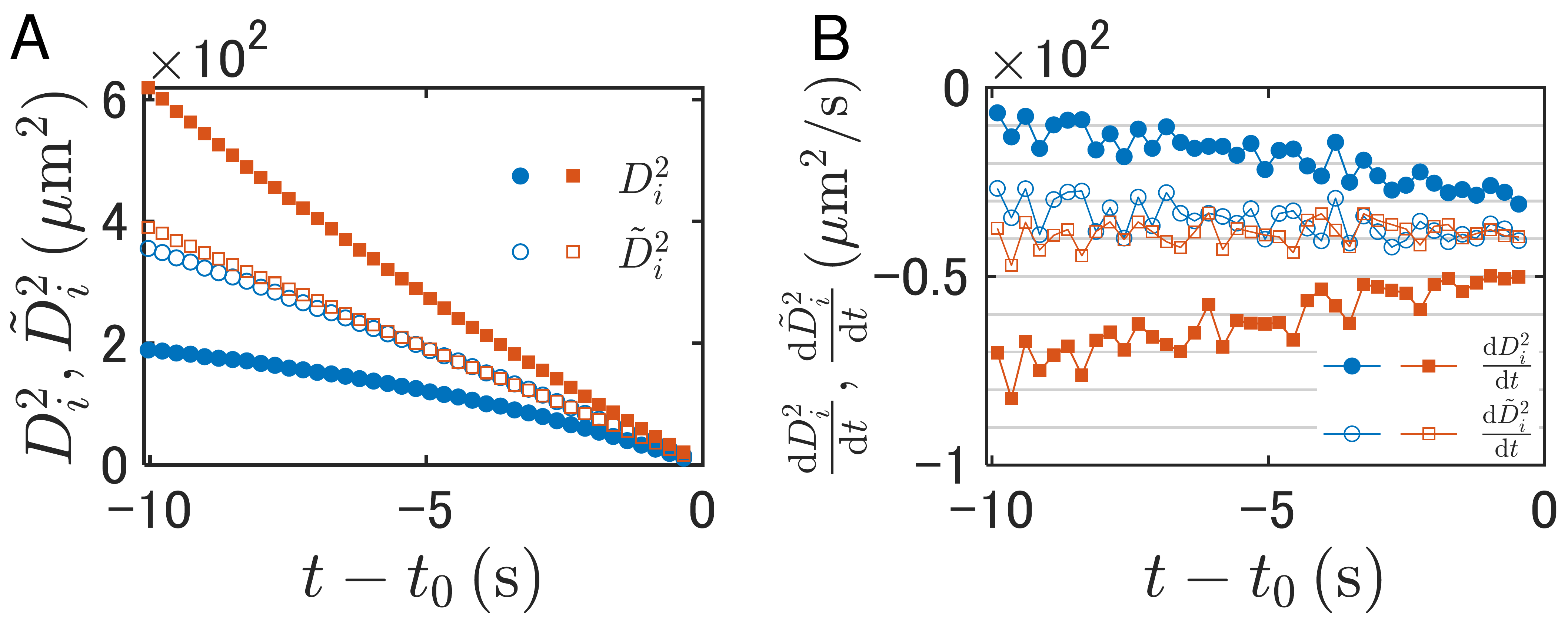}
\centering
\caption{Comparison between the distance $D_i$ in the laboratory frame and $\tilde{D}_i$ in the co-moving frame. (\textit{A}) The squared distances $D_i^2$ and $\tilde{D}_i^2$ against time $t-t_0$, for the reconnection event shown in \figref{M-fig3}C,E. (\textit{B}) The derivative of the data shown in (\textit{A}). The results in the co-moving frame, $\diff{\tilde{D}_i^2}{t}$, appear horizontal in this plot, indicating the validity of the scaling law $\tilde{D}_i \propto |t-t_0|^{1/2}$. Note that this scaling law is seen only in a much narrower range near $t \approx t_0$ for $D_i$ in the laboratory frame.}
\label{figS:CompDist}
\end{figure}

\begin{figure}
\includegraphics[width=0.9\hsize,clip]{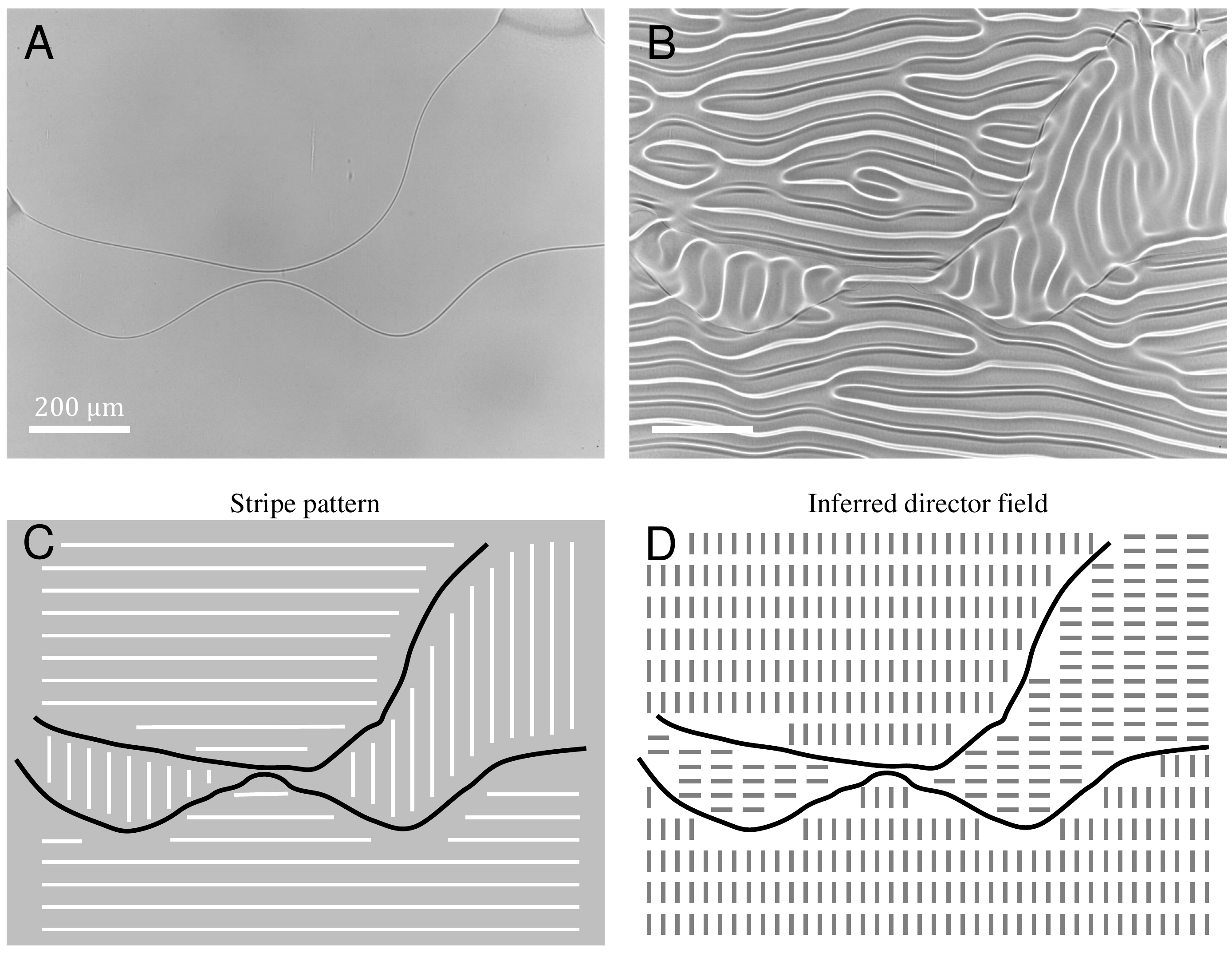}
\centering
\caption{Observation of the roll convection around a pair of disclinations during the relaxation process. The disclinations were about to reconnect, but a voltage of $14\unit{V}$ RMS was applied before the reconnection, to induce the roll convection (Williams domain). (\textit{A},\textit{B}) The images before and after the application of the voltage, respectively. The voltage was applied roughly $0.5\unit{s}$ after the image (\textit{A}), and about $2.0\unit{s}$ later the image (\textit{B}) was captured. (\textit{C}) Sketch of the stripe pattern observed in (\textit{B}). (\textit{D}) Sketch of the director field in the midplane, inferred from this observation. This indicates the twist configuration, as sketched in \figref{M-fig4}B bottom.}
\label{figS:WD}
\end{figure}

\begin{figure}
\includegraphics[width=0.8\hsize,clip]{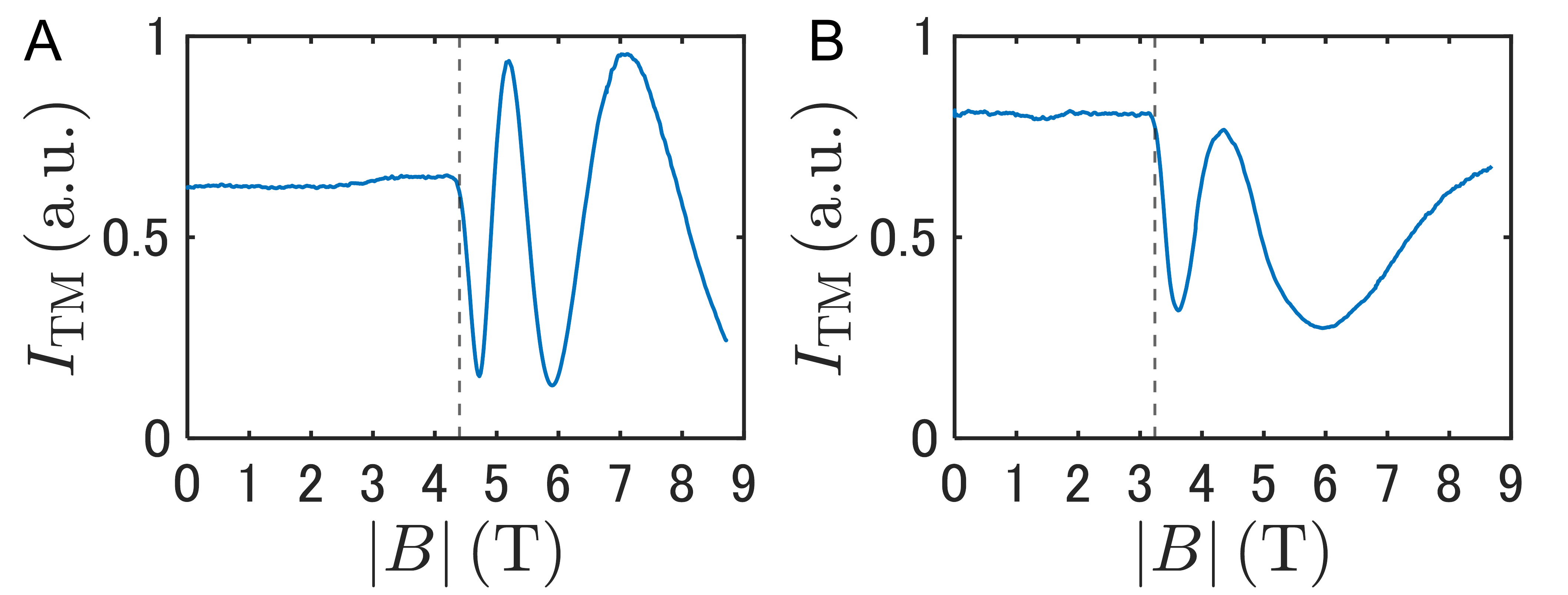}
\centering
\caption{Observation of the Fr\'eedericksz transition of MLC-2037, for the splay (\textit{A}) and twist (\textit{B}) configurations. The intensity of the transmitted light $I_\mathrm{TM}$ is shown against the applied magnetic flux density $B$. The data shown here were obtained for increasing $B$ and by using a halogen lamp as a light source. The dashed lines indicate $B^\mathrm{c}_i$ from these data.}
\label{figS:K2}
\end{figure}

\begin{figure}
\includegraphics[width=\hsize,clip]{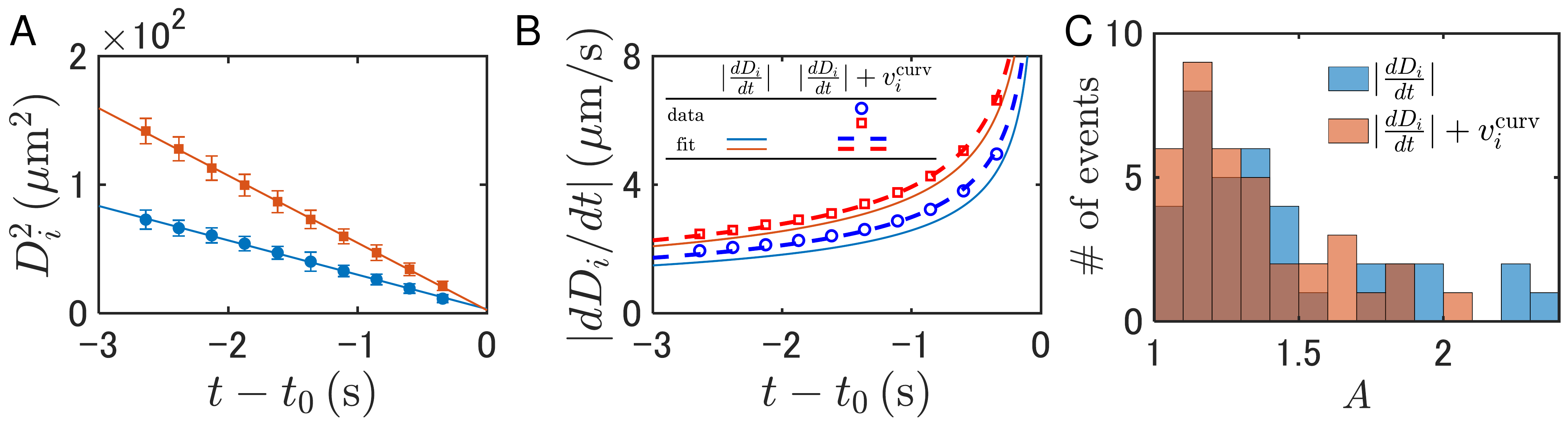}
\centering
\caption{Evaluation of curvature effect. (\textit{A}) Distance $D_i(t)$ measured in the laboratory frame, for an example pair of reconnecting disclinations. This panel is identical to \figref{M-fig3}C. (\textit{B}) Speed $\left|\diff{D_i}{t}\right|$ for the disclination pair shown in (\textit{A}). The solid lines indicate the speed evaluated from the fits shown in (\textit{A}), without compensation of the curvature effect. The results after compensation of the curvature effect (see text) are shown by open symbols (data) and dashed lines (fit by \eqref{eq:speed_curv}). (\textit{C}) Histograms of the asymmetry parameter $A$, with (red) and without (blue) the compensation of the curvature effect. Note that the three outliers described in the caption of \figref{M-fig3}E are also outside the range displayed here, for both of the histograms ($A \approx 3.2, 5.5, 40$ become $A \approx 2.4, 3.7, 8.6$, respectively, by considering the curvature effect).}
\label{figS:curv}
\end{figure}



\FloatBarrier

\movie{
A relaxation process from the electrohydrodynamic turbulence upon removal of the applied voltage. Real playback speed. The spacing between $z$ slices was $1.5\unit{\mu m}$. The number of voxels was $512, 512, 20$ in the $x, y, z$ directions, respectively. The time interval between consecutive confocal images was $0.760\unit{s}$. The other conditions were the same as those for the in-plane reconnections.
}

\movie{
The in-plane reconnection shown in \figref{M-fig1}D. Real playback speed. 
}

\movie{
The intersecting reconnection shown in \figref{M-fig1}E. Real playback speed. 
}

\movie{
The loop shrinkage shown in \figref{M-fig1}F. Real playback speed.
}



\bibliography{ref}